\documentclass[doublecol]{epl2} 
\newcommand{\bea}{\begin{eqnarray}}    
\newcommand{\eea}{\end{eqnarray}}      
\newcommand{\be}{\begin{equation}}
\newcommand{\ee}{\end{equation}}
\newcommand{\bef}{\begin{figure}}
\newcommand{\eef}{\end{figure}}

\def\spose#1{\hbox to 0pt{#1\hss}}
\def\ltapprox{\mathrel{\spose{\lower 3pt\hbox{$\mathchar"218$}}
\raise 2.0pt\hbox{$\mathchar"13C$}}}
\def\gtapprox{\mathrel{\spose{\lower 3pt\hbox{$\mathchar"218$}}
\raise 2.0pt\hbox{$\mathchar"13E$}}}
\def\inapprox{\mathrel{\spose{\lower 3pt\hbox{$\mathchar"218$}}
\raise 2.0pt\hbox{$\mathchar"232$}}}

\title{Absence of self-averaging and of homogeneity in the 
large scale galaxy distribution}
\shorttitle{Inhomogeneity of galaxy distribution} 

\author{Francesco Sylos Labini\inst{1,2} 
\and Nikolay L. Vasilyev \inst{3} 
\and Luciano Pietronero \inst{2,4}
\and Yurij V. Baryshev \inst{3}
}
\shortauthor{F. Sylos Labini \etal}

\institute{                    
  \inst{1} Museo Storico della Fisica e Centro Studi e
  Ricerche Enrico Fermi, - Piazzale del Viminale 1,
  00184 Rome, Italy \\
  \inst{2}  Istituto dei Sistemi 
  Complessi CNR, - Via dei Taurini 19, 00185 Rome, Italy \\
 \inst{3} Institute of Astronomy, St.Petersburg State
  University - Staryj Peterhoff, 198504, St.Petersburg, Russia\\
 \inst{4} Dipartimento di Fisica, Universit\`a di Roma
  ``Sapienza'' - P.le Aldo Moro 2, 00185, Rome Italy
}
\pacs{98.80.-k}{Cosmology}
\pacs{05.40.-a}{Fluctuations phenomena in random processes}
\pacs{02.50.-r}{02.50.-r Probability theory, stochastic processes, 
and statistics}

\abstract{ The properties of the galaxy distribution at large scales
  are usually studied using statistics which are assumed to be
  self-averaging inside a given sample.  We present a new analysis
  able to quantitatively map galaxy large scale structures while
  testing for the stability of average statistical quantities in
  different sample regions.  We find that the newest samples of the
  Sloan Digital Sky Survey provide unambiguous evidence that galaxy
  structures correspond to large amplitude density fluctuations at all
  scales limited only by sample sizes.  The two-point correlations
  properties are self-averaging up to approximately $30$ Mpc/h and are
  characterized by a fractal dimension $D=2.1\pm0.1$.  Then at all
  larger scales probed density fluctuations are too large in amplitude
  and too extended in space to be self-averaging inside the considered
  volumes.  These inhomogeneities are compatible with a continuation
  of fractal correlations but incompatible with: (i) a homogeneity
  scale smaller than 100 Mpc/h, (ii) predictions of standard
  theoretical models, (iii) mock galaxy catalogs generated from
  cosmological Nbody simuations.}

\begin{document}

\maketitle

\section{Introduction}

Understanding the large scale structure of the universe as mapped by
galaxy distribution represents one of the cornerstones of modern
cosmology. It provides the basic test for theories of structure
formation in the universe.  A primary question in the statistical
analysis of three-dimensional galaxy catalogs (where, in addition to
the angular coordinates, the redshift is measured  and through
Hubble's law \cite{pee80} the distance of each object) concerns the
determination of a scale where the distribution becomes
homogeneous. Such a scale $\lambda_0$ can be defined to be the one
beyond which counts of galaxies in three dimensional spherical volumes
of radius $r$ grow as $ r^3$ \cite{book}.  


A decade ago, by measuring the conditional density, i.e., the local
galaxy density seen by a galaxy in a spherical volume of radius $r$
around itself \cite{pie87,book}, some of us found that galaxy
correlations are power-law with an exponent $\gamma \approx 1$ up to
the sample sizes, i.e., $\sim 30$ Mpc/h
\footnote{We use $H_0=100h$ km/sec/Mpc, with $0.4\le h \le 0.7$, for
  the Hubble's constant}, corresponding to a fractal dimension
$D=3-\gamma\approx 2$ \cite{cp92,slmp98,jmsl99}.  These results were
in contrast with the analysis of the same samples by, e.g.,
\cite{dp83,park,benoist}, who found $\lambda_0 \approx 10$ Mpc/h and
$\gamma=1.8$. The reason for these differences lies in the a-priori
assumption of homogeneity, inside a given sample, of the standard
statistical analysis \cite{pie87,slmp98,book}.

At larger scales, with weaker statistical significance, there was an
evidence compatible with the fact that power-law correlations in the
conditional density extend up to $r \sim 100$ Mpc/h or more
\cite{slmp98,esp}.  These results generated a debate in the field
\cite{wu,davis,pmsl96} because even though galaxy structures were
found in many different catalogs to extend to scales of the order of
hundreds of megaparsecs, the characteristic length scale $\lambda_0$
statistically describing their correlations was determined to be a few
megaparsecs \cite{dp83,park,benoist}.  While for some this was a
paradox \cite{pie87,cp92,slmp98}, for others
\cite{dp83,park,davis,benoist,wu} the explanation was that large scale
structures have small amplitude relative to the average density.
However this interpretation is problematic as in the range of scales
where the conditional density shows power-law correlations the sample
density is not well defined while density fluctuations have large
amplitude \cite{pie87,slmp98,book}.  The determination of the
crossover scale $\lambda_0$, where the conditional density from a
power-law turns to a constant, allowing a meaningful determination of
the average density, has been thus an important task of galaxy
correlations studies in the last decade \cite{bt05}.

Two new galaxy catalogs, the Sloan Digital Sky Survey (SDSS ---
\cite{york}) and the Two degree Field Galaxy Redshift Survey
\cite{colless03}, have recently provided great advances in the mapping
of the local universe both for the number of objects measured in
continuously growing volumes and for the determination of several
parameters for each of them.  Several studies
\cite{2df_paper,dr4_paper,hogg,tikonov} of different samples of these
surveys confirmed the small scale correlations measured by
\cite{slmp98,jmsl99}.  In addition it has been claimed that a slow
crossover toward homogeneity occurs \cite{hogg} with the average
conditional density in spheres at $\sim 20$ Mpc/h having twice the
amplitude of the asymptotic density reached at $r>70$ Mpc/h
\cite{tikonov}.  It was however noticed that galaxy structures could
bias the determination of correlations in these samples introducing
uncontrolled systematic effects \cite{joyce05,dr4_paper,2df_paper}.
Recently in the Two degree Field Galaxy Redshift Survey it has been
found that \cite{paper_2df_prl,paper_2df_aea} galaxy distribution
characterized by large amplitude fluctuations with a large spatial
extension, whose size is only limited by the sample's boundaries. In
addition at scales $r < 40$ Mpc/h, it has been observed a well defined
and statistically stable power-law behavior of the average number of
galaxies in spheres in agreement with previous determinations.

A different and complementary method to characterize structures is
provided by galaxy counts as a function of the radial distance from
us or of the apparent luminosity \cite{gsl00}. These show large
fluctuations around the average behavior both in redshift
\cite{kerscher98} and angular surveys \cite{picard91,bd97}.  There
have been controversies as to whether these are due to real clustering
or to incompleteness of the catalogs \cite{wu,gsl00}.  Recent results
support the conclusion that the local galaxy distribution is
characterized by large scale structures with significant correlations
on scales $r > 50$ Mpc/h \cite{busswell03,frith03}.

In this paper we use a new method which is able to establish, in a
given sample, the link between the small scales $r<30$ Mpc/h
correlations and the large scales $r >30$ Mpc/h fluctuations in galaxy
counts and which clarifies how the latter influence the determination
of the former. Using it we can test whether sample means, variances
and correlations are well defined, i.e., whether they are
statistically stable in different sub-volumes of the given sample.  By
applying this method to the data of the SDSS project \cite{paperdr6}
we detect large density fluctuations of spatial extension limited by
the samples' sizes. We show that these introduce systematic biases in
the determination of large scale correlations.

\section{The Data}
\label{samples_sdss}

The SDSS \cite{york} is currently the largest spectroscopic survey of
extragalactic. Here we consider 
the main galaxy (MG) sample (DR6) \cite{paperdr6}
containing redshifts for about 800,000 galaxies which 
covers an area of 7425 square degrees on the celestial sphere.
To query the DR6 database we constrain the flags indicating the type
of object so that we select only the galaxies from the MG sample.  We
then consider galaxies in the redshift range $10^{-4} \leq z \leq
0.3$. The redshift confidence parameter is constrained to be $z_{conf}
\ge 0.35$ with flags indicating no significant redshift determination
errors.  In addition we apply the filtering condition $m_r < 17.77$,
using Petrosian apparent magnitudes in the $r$ filter which are
corrected for galactic absorption, and thus taking into account the
target magnitude limit for the MG sample in the DR6
\cite{strauss2002}.  In this way we have selected 479,417 objects.  We
considered also more stringent limits in apparent magnitude, to test
whether a possible incompleteness of the survey at bright and/or faint
apparent magnitudes could generate a fake signal. 
To this aim we used 
we have 
$14.5 \le m_r \le 17.5$
and we selected 370,893 objects, i.e., about $25 \%$ less than with
less conservative constraints.  We have considered a rectangular
angular fields, with uniform coverage, in the SDSS internal angular
coordinates $(\eta,\lambda)$ limited by $-6^{\circ} \le \eta \le
36^\circ$ and $-48^\circ \le \lambda \le 32.5^\circ$.

%
To construct volume limited (VL) samples that are unbiased for the
selection effect related to the cuts in the apparent magnitude, we
have applied a standard procedure \cite{zehavietal02}.  Firstly we
compute metric distances $R(z)$ using the standard cosmological
parameters $\Omega_M=0.3$ and $\Omega_\Lambda=0.7$.  
Secondly the
galaxy absolute magnitude is determined to be $ M_r = m_r - 5 \cdot
\log_{10}\left[R(z) \cdot (1+z)\right] - K_r(z) - 25\;, $ where
$K_r(z)$ is the K-correction.  We determine the $K_r(z)$ term from the
NYU VACG data \cite{nyu_vacg}.  Finally, we have considered two
different VL samples (named VL1 and VL2) defined by two chosen limits
in absolute magnitude and metric distance: for VL1 $R\in [100,300]$
Mpc/h, $M\in [-22,-20]$ and for VL2 $R\in[200,600]$ Mpc/h,
$M\in[-23,-21.5]$.  The number of galaxies is about $4\cdot 10^4$ in
VL1 and $3 \cdot 10^4$ in VL2.  Different cuts in absolute magnitude
do not introduce substantial differences in the results presented in
this paper.  When more conservative and stringent limits in apparent
magnitude are applied we find that in the samples with the same limits
in distance, there are up to three times less galaxies. However the
main results presented in this paper are affected only in the fact
that statistics is less robust.

The Millennium project \cite{springel05} has performed several
cosmological simulations of standard theoretical models.  Amount of
dark matter and cosmological parameters are given in agreement with
standard models.  The dark matter simulations have about $10^{10}$
particles. From these galaxies are identified according to
semi-analytics models of galaxy formation \cite{cronton06}.  We have
cut a sample with exactly the same geometry as the SDSS VL1 sample and
a sample close to the geometrical parameters of the SDSS VL2 applying
the same absolute magnitude limits in $r$-filter as for the SDSS data.
In the SDSS we use a redshift space analysis while in mock catalogs a
real space one.  The difference between the real and redshift space
analysis is relevant for very small scales, i.e., $r<5$ Mpc/h and thus
does not influence results on scales of the order of $100$ Mpc/h
\cite{bt05,paper_2df_prl,paper_2df_aea}.

\section{Statistical methods}

Statistical properties are determined by making averages over the
whole sample volume~\cite{book}.  In doing so one implicitly assumes
that a certain quantity  measured in different regions of the sample
is statistically stable, i.e., that fluctuations in different
sub-regions are described by the same probability density function
(PDF).  However it may happen that measurements in different
sub-regions show systematic   differences, which
depend, for instance, on the spatial position of the specific
sub-regions.  In this case the considered statistic is not
statistically stationary in space, the PDF systematically differs in
different sub-regions and its whole-sample average value is 
not a meaningful descriptor \cite{book}.

In general such systematic differences may be related to two different
possibilities: (i) that the underlying distribution is not
translationally and/or rotationally invariant; (ii) that the volumes
considered are not large enough for fluctuations to be self-averaging
\cite{Aharony}.
On very general grounds, we expect the galaxy distribution to satisfy
the condition of {\it statistical} stationarity in space to avoid
special points or directions \cite{pee80,book}.  Hence the question we
face in a finite volume analysis concerns whether it is large enough
to obtain statistically stable results. Note that stationary
stochastic distributions satisfy the condition of spatial statistical
isotropy and homogeneity also when they have zero average density in
the infinite volume limit \cite{book}. This condition is called the
Conditional Cosmological Principle \cite{book} to differentiate it
from the stronger Cosmological Principle which requires exact
homogeneity and deterministic rotational and translation invariance
\cite{pee80,book}.

For the case of galaxy surveys there is an intrinsic preferred
direction which is set by the radial position from the observer, i.e.,
the Earth.  It is thus necessary to show that statistical quantities
do not depend on the radial distance from us.  To this aim, in a given
sample, a simple approach is to determine the number $N(r;R)$ of
galaxies in spheres of radius $r$, centered on a galaxy whose distance
from the origin is $R$: we call it the scale-length (SL) analysis.  As
we discuss below, this is found to be very efficient in mapping large
scale structures which appear as large fluctuations of $N(r;R)$.  For
instance by studying it in various angular slices of the SDSS samples
we identify a giant filament covering, in the largest contiguous
angular area of the survey, more than 400 Mpc/h at $R\sim 500$ Mpc/h.
In different sky directions the SL analysis reveals a variety of
structures, showing that large density fluctuations are quite typical.

Averaged over the whole sample the quantity $N(r;R)$ gives an estimate
of the average conditional number of galaxies in spheres of radius
$r$.  An estimator making the weakest a-priori-assumptions about the 
properties of the distribution outside the sample volume is 
\cite{pie87,slmp98,book} 
\be
\label{eq1} 
\overline{N(r)} = \frac{1}{M(r)} \sum_{i=1}^{M(r)} N_i(r) \;, \ee
where $N_i(r)$ is the number of galaxies seen by the $i^{th}$
center-point and the number of centers $M(r)$ varies with $r$ because
only those galaxies for which the sphere is fully included in the
sample volume are considered as centers \cite{book}. Even in this case,
there is an intrinsic selection effect related to the geometry of the
samples, which are portions of spheres: when $r$ is large only a part
of the sample is explored by the volume average. Hence for large
sphere radii $M(r)$ decreases and the location of the galaxies
contributing to the average in Eq.\ref{eq1} is mostly at radial
distance $\sim [R_{min}+r$, $R_{max}-r]$ from the radial boundaries of
the sample at $[R_{min}$, $R_{max}]$.

When Eq.\ref{eq1} scales as $ \overline{N(r)} \sim r^D$ and $D=3$ the
distribution is homogeneous, while for $D<3$ it is fractal
\cite{gsl00,book}.  Furthermore fluctuations
$
 \delta^2(r) = [\overline{N(r)^2} - \overline{N(r)}^2] / \overline{N(r)}^2
$
are small for a homogeneous distribution with any kind of
small-amplitude correlations ($\delta^2(r) \ll 1$) and large for a
fractal one ($\delta^2(r) \sim 1$) \cite{gsl00,book}.  To study
fluctuations we determine the PDF of $N_i(r)$, which is expected to
converge to a Gaussian when $r \gg \lambda_0$ \cite{book}.

\section{Results} 
\begin{figure}
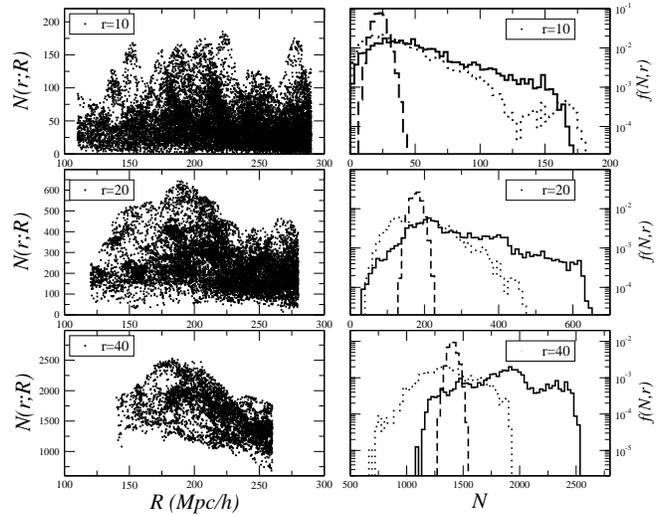

\onefigure[scale=0.32]{Fig1.eps}
\caption{{\it Left panels}: From top to bottom the SL analysis for the
  SDSS sample VL1, with $r=10,20,40$ Mpc/h.  
{\it Right panels}: Probability density
  function of $N(r;R)$ in two non-overlapping sub-samples with equal
  volume (each half of the sample volume) at small and large $R$.
  While for $r=10,20$ Mpc/h the PDF (nearby sub-sample solid line and
  faraway sub-sample dotted line) is reasonably statistically stable,
  for $r=40$ Mpc/h there is a clear difference.  The dashed line
  corresponds to the Poisson distribution: a Gaussian function gives
  very good fits for all $r$.  }
\label{figure2}
\end{figure}

Let us now consider the VL1 sample. Here the SL analysis (Fig.\ref{figure2})
detects large density fluctuations without a clear radial-distance
dependent trend.  Correspondingly the PDF has a regular shape
characterized by a peak with a long $N$ tail and it is sufficiently
statistically stable in different non-overlapping sub-samples of equal
volume.  This occurs except for the largest sphere radii, i.e., for
$r>$30 Mpc/h, for which the number of independent centers becomes too
small.

\begin{figure}
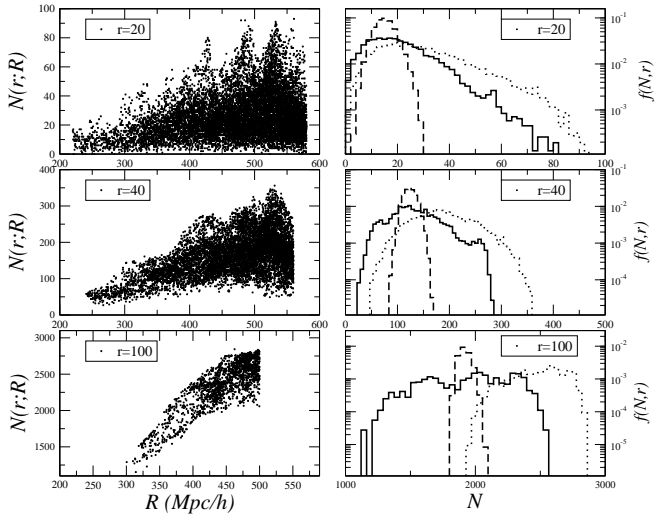

\onefigure[scale=0.32]{Fig2.eps}
\caption{
The same of Fig.\ref{figure2} for the VL2 sample 
with $r=20,40,100$ Mpc/h.   The fact that the signal for SDSS data
  becomes smoother when $r$ increases, but still with a systematic
  radial distance-trend, is due to the fact that many spheres overlap
  when $r$ growths.    While for $r=20$ Mpc/h the PDF 
   is reasonably statistically stable, for $r=40,100$ Mpc/h
  there is a clear difference.  }
\label{figure3}
\end{figure}

In the deeper VL2 sample we find instead a systematic trend of $N(r;R)
$ as a function of $R$ (Fig.\ref{figure3}).
Particularly $N(r;R)$, for $R>300$ Mpc/h, grows without any clear
saturation, for sphere radii up to $r \sim 100$ Mpc/h.  The PDF
in two non-overlapping sub-samples of equal volume is found to differ
systematically for $r>30$ Mpc/h, and its average value moves as a
function of $R$.  This shows that fluctuations are not self-averaging
at those scales.  This behavior is due to the large scale structures
at scales $R> 300$ Mpc/h.  However at smaller scales, i.e., $r<30$
Mpc/h, the PDF in different sub-samples is reasonably statistically
stable and similar to the one found in VL1.  This shows that, at those
scales, fluctuations are self-averaging because the volume average can
explore different regions of the sample.  Thus these results show
that, at the largest scales probed, there are large density
fluctuations which are not self-averaging because of the limited
sample volume. These determine relative variations larger than unity
in the estimation of the average density in spheres of radius $r=100$
Mpc/h.  We thus conclude that the homogeneity scale must be $\lambda_0
> 100$ Mpc/h, the largest sphere radius we considered.

\begin{figure}
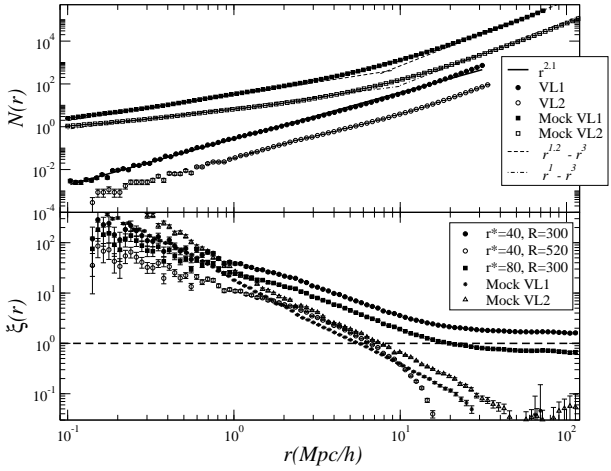

\onefigure[scale=0.32]{Fig3.eps}
\caption{{\it Upper Panel}: The sample average conditional number of
  galaxies (Eq.\ref{eq1}) for the SDSS VL1 and VL2 samples up to
  $r<30$ Mpc/h.  The best fit slope (soldi line) gives $D= 2.1 \pm
  0.1$.  The difference in amplitude between the two samples is simply
  ascribed to a luminosity selection effect \cite{book}, as VL1
  contains fainter galaxies than VL2.  The amplitude of the mock VL1
  and VL2 samples has been rescaled by the same arbitrary factor for
  seek of clarity.  {\it Bottom Panel}: Standard two-point correlation
  function in the VL2 sample estimated by Eq.\ref{xi2}: the sample
  average density is computed in spheres of radius $r^*$ and
  considering all center-points lying in a bin of thickness $\Delta
  R=40$ Mpc/h centered at different radial distance $R$. The case
  $r^*=80$ Mpc/h and $R=520$ Mpc/h gives the average over the whole
  sample, i.e., Eq.\ref{xi}, and it coincides with the estimation of
  \cite{zehavietal02} in a similar sample.  The flat tail of $\xi(r)$,
  for $r^*=40,80$ Mpc/h and $R=300$ Mpc/h, reflects the
  inhomogeneities with strong correlations at large scales.  The stars
  and triangles correspond to the behavior of the mock VL1 and VL2
  samples: in this case the amplitude is statistically stable and thus
  meaningful.  The small amplitude difference in this case is ascribed
  to the different selection in luminosity \cite{springel05}.  }
\label{figure4}
\end{figure}

Previous analyses of smaller galaxy catalogs, e.g.,  
\cite{cp92,slmp98,jmsl99,dr4_paper,2df_paper,tikonov,
  zehavietal02,norberg02,hogg}, considered
sample averaged statistics without quantitatively testing whether a
significant bias could affect the results (but see
\cite{dr4_paper,2df_paper}).  For instance the estimator of the most
commonly used statistics, the two-point correlation function
\cite{tk69}, can be written as \cite{book} 
\be
\label{xi} 
\xi(r) +1 = \frac{\overline{N(r,\Delta r)}} {V(r,\Delta r)} \cdot
\frac{V} {N} \;.  
\ee 
The first ratio in the r.h.s. of Eq.\ref{xi} is
the average  conditional density, i.e., the number of galaxies in shells
of thickness $\Delta r$ averaged over the whole sample, divided by the
volume $V(r,\Delta r)$ of the shell. The second ratio in the r.h.s. of
Eq.\ref{xi} is the average density estimated in a sample containing $N$
galaxies and with volume $V$.  When measuring this function we
implicitly assume, in a given sample, that: (i) fluctuations are
self-averaging in different sub-volumes \cite{book} (ii) the linear
dimension of the sample volume is $V^{1/3} \gg \lambda_0$
\cite{pie87,book}, i.e., the distribution has reached homogeneity
inside the sample volume.  When the latter condition is not verified
the $\xi(r)$ analysis is biased by
systematic finite size effects even if 
fluctuations are self-averaging
\cite{pie87,book}.
To show how non self-averaging fluctuations inside a given sample
bias the $\xi(r)$ analysis, we consider the estimator
\be 
\label{xi2}
\xi(r;R,\Delta R) +1 = \frac{\overline{N(r,\Delta r)}} {V(r, \Delta
  r)} \cdot \frac{V(r^*)}{\overline{N(r^*;R,\Delta R)}} \,, 
  \ee where
the second ratio on the r.h.s. is the   density of points
in spheres of radius $r^*$ averaged over  the centers
lying  in a shell of thickness $\Delta R$
around the radial distance $R$.  If the distribution is homogeneous,
i.e., $r^*>\lambda_0$, and statistically stationary, Eq.\ref{xi2}
should be statistically independent on the range of radial distances
$(R,\Delta R)$ considered.  For instance we consider, in the VL2 sample, $\Delta
R=40$ Mpc/h and $R = 240$ Mpc/h or $R=520$ Mpc/h, with $r^* > 50$
Mpc/h. We thus find large variations in the amplitude of $\xi(r)$
(Fig.\ref{figure4}).  This is simply an artifact generated by the
large density fluctuations on scales of the order of the sample sizes.  The
results that the estimator Eq.\ref{xi} (or others based on
pair-counting \cite{book,cdm_theo}) has nearly the same amplitude in
different samples,
e.g.,
 \cite{dp83,park,benoist,zehavietal02,norberg02},
despite the large fluctuations of $N(r;R)$, are 
simply explained by the fact that $\xi(r)$ is a ratio between the
local conditional density and the sample average density: 
both vary in the same way when the radial distance is
changed and thus the amplitude is nearly constant.

On the other hand Eq.\ref{eq1}, {\it averaged over the volumes where
  the PDF has a statistically stable shape}, shows in both considered
samples a power-law behavior for $r<$ 30 Mpc/h corresponding to a
fractal dimension $D= 2.1 \pm 0.1$ in agreement with
\cite{slmp98,jmsl99,hogg,2df_paper,dr4_paper} (Fig.\ref{figure4}).
Due to the non self-averaging nature of fluctuations at larger scales,
i.e., due to limited volumes, we are not able to determine
correlations for $r>30$ Mpc/h.

\section{Discussion}  
According to standard models of cosmological structure formation,
gravitational clustering gives rise to non-linear perturbations from
homogeneous initial conditions in the early universe \cite{pee80}. If
the initial amplitude of fluctuations is normalized to the
anisotropies of the Cosmic Microwave Background Radiation (CMBR)
\cite{spergel}, then the homogeneity scale is about $\lambda_0^m = 10$
Mpc/h \cite{joyce05}, i.e., twice the value at which $\xi(r)=1$
\cite{pee80,book}.

\begin{figure}
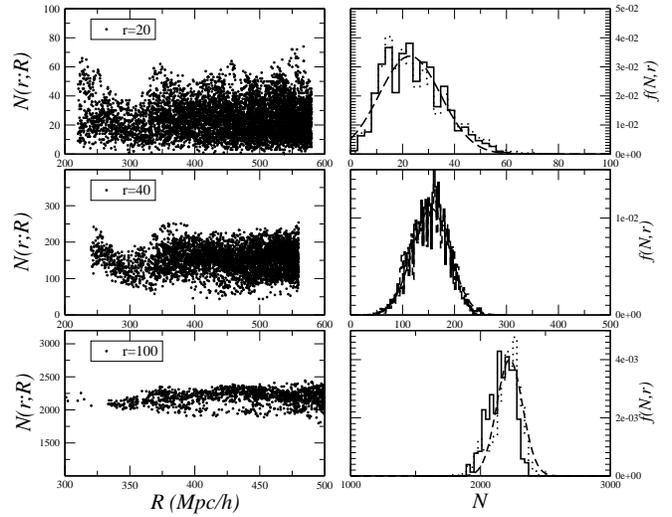

\onefigure[scale=0.32]{Fig4.eps}
\caption{ The same of fig.\ref{figure2} for mock sample VL2, with
  $r=20,40,100$ Mpc/h.  Although for $r=20$ Mpc/h fluctuations are
  still important, they rapidly become small for larger radii.  The
  PDF is statistically stable for all $r$ and $R$.  The solid line
  corresponds to a Gaussian fit.  }
\label{figure5}
\end{figure}

Indeed in mock galaxy catalogs generated from N-body simulations of
standard cosmological models \cite{springel05,cronton06}, $N(r;R)$
does not show, for $r >\lambda_0^m$, large fluctuations or systematic
trends as a function of $R$ (Fig.\ref{figure5}).  
Because in these
artificial catalogs fluctuations are small and self-averaging,
whole-sample averaged statistics are meaningful at all scales. From
the $\overline{N(r)}$ analysis we find, differently from the real
galaxy data, that $\overline{N(r)} \sim r^{D}$ with $D=1.1 \pm 0.1$
for $r<\lambda_0^m$ and $D=3$ for $r>\lambda_0^m$ (Fig.\ref{figure4}).
Correspondingly the PDF rapidly converges to a Gaussian for
$r>\lambda_0^m$.  The $\xi(r)$ function, estimated by Eq.\ref{xi} or
Eq.\ref{xi2}, has a statistically stable amplitude. The residual small
amplitude difference between the mock VL1 and VL2 samples is ascribed
to a different selection in luminosity \cite{springel05}.  The
amplitude of $\xi(r)$ is equal to unity at $r \approx 6 $ Mpc/h in
agreement with \cite{springel05,cronton06}. In addition its shape
reasonably agrees with the standard estimation of $\xi(r)$ from galaxy
catalogs
\cite{dp83,park,benoist,zehavietal02,norberg02}.
However, as discussed above, the latter estimation is biased by
systematic effects making the agreement fortuitous.

Standard theoretical models predict that, for $r>\lambda_0^m$, the
small fluctuations in the early universe are linearly amplified by
gravitational clustering. Therefore for $r>\lambda_0^{m}$ the shape of
the theoretical $\xi^m(r)$ must be the same as the initial one
\cite{pee80}.  This is characterized by a length-scale $r_c$, where
$\xi^m(r_c)=0$, which is fixed by the physics of the early universe
and estimated from CMBR anisotropies to be $r_c \approx$ 100 Mpc/h
\cite{spergel,cdm_theo}.  For $r>r_c$, $\xi^m(r)$ becomes negative,
corresponding to super-homogeneous correlations characterized by the
most rapid possible decay of fluctuations \cite{glass,torquato}.  This
theoretical framework applies to the whole mass distribution, where
dark matter is supposed to provide with the main contribution.
Galaxies would form on the largest peaks of the density
field. Standard models of galaxy formation describe this physical
phenomenon as a selection mechanism \cite{kaiser}.  This leaves
unperturbed the scale $r_c$ and slightly changes $\lambda_0^m$
\cite{bias,cdm_theo}.  We find $\lambda_0 > 100 $ Mpc/h $\ge
r_c \gg \lambda_0^{m}$.  This raises a fundamental inconsistency for
the relation between galaxy structures and CMBR anisotropies as no
physical mechanism is known, which by sampling a super-homogeneous
density field transforms it into a strongly inhomogeneous one
\cite{bias,book,cdm_theo}.

\section{Conclusion} 

In summary, by applying the SL analysis to the newest SDSS galaxy
samples, we measure large density fluctuations of spatial extension
limited by sample sizes.  At scales $r<30$ Mpc/h we detect
statistically stable fractal correlations with $D=2.1 \pm 0.1$.  On
larger scales, $r>30$ Mpc/h, we find that the galaxy distribution is
strongly inhomogeneous and fluctuations are not self-averaging in the
samples considered.  This situation is compatible with fractal
power-law correlations extending to such length-scales but 
incompatible with homogeneity at $\lambda_0 \le 100$ Mpc/h.  Indeed,
in a portion of a fractal, large structures are expected to be present
at any scale, fluctuations being self-averaging only if the sample
volume is large enough \cite{book}.
These results have important consequences on the theoretical
interpretation of the large scale universe, where models, normalized
to CMBR anisotropies, predict there is not enough time to form
structures with relative density fluctuations larger than unity on
scales larger than $\lambda_0^m \approx 10$Mpc/h
\cite{pee80,springel05}.  This length scale is more than ten times
smaller than our lower limit to $\lambda_0$. Indeed the latter is of
the order of the scale $r_c$ where theoretical model predict matter
distribution to have negative correlations, a situation which is in
contrast with the results from the data analyzed here.  Thus the large
scale inhomogeneities detected in the SDSS samples are incompatible
with the predictions of standard theoretical models relating the early
universe physics, with CMBR normalization, to structures in the
present universe.  Moreover we found that for $r<\lambda_0^m$, mock
galaxy catalogs have different correlations from real galaxy data,
i.e., $D = 1.1 \pm 0.1$ instead of $D=2.1 \pm 0.1$.  Thus structures
generated by N-body simulations are intrinsically different from
observed ones.

Recently three dimensional maps of dark matter distribution from
weak lensing observations have been published \cite{massey}.
Dark matter is observed to trace the same structures as
galaxies. Thus the whole matter
distribution would be inhomogeneous on scales larger than $100$ Mpc/h.
This has a great impact on the whole
theoretical framework for the physical understanding of the large
scale universe.  For instance it may imply a new type of evolution
scenario within an open Friedmann model \cite{pwa} or new types of
spatial averaging of the Einstein equations \cite{buchert,wiltshire}
which relate the observed inhomogeneities to the apparent acceleration
measured from supernovae observations \cite{perl}.

The determination of statistical properties 
of those very large structures, which we detected but could not analyze 
in detail,  should be
possible when sample volumes  become large enough so that the
corresponding fluctuations will be self-averaging on scales larger
than the ones studied here.  The application of the SL analysis to the
forthcoming galaxy redshift surveys, like the complete SDSS
\cite{york}, thus represents  an important  task.

\acknowledgments We warmly thank M. Joyce for his careful reading of
the manuscript and the very detailed comments and suggestions.  We
thank A. Gabrielli, R. Durrer, M. Lopez Correidoira for useful remarks
and discussions. We acknowledge the use of the Sloan Digital Sky
Survey data \cite{paperdr6} and of the Millennium run semi-analytic
galaxy catalog \cite{cronton06}

\end{document}